\documentclass[]{aa}
\usepackage{graphicx}
\usepackage[authoryear]{natbib}
\usepackage{tabls}
\bibpunct{(}{)}{;}{a}{}{,}
\usepackage{txfonts}
\begin{document}
\title{INTEGRAL long-term monitoring of the Supergiant Fast X-ray Transient XTE~J1739$-$302}


\author{ P. Blay\inst{1}
\and S. Mart\'{\i}nez-N\'u\~{n}ez\inst{2,1}
\and
I.~Negueruela\inst{2}
\and
K.~Pottschmidt\inst{3,4}
\and
 D.~M.~Smith\inst{5}
\and
J.~M.~Torrej\'on\inst{2,6}
\and
P.~Reig\inst{7,8}
\and
P.~Kretschmar\inst{9}
\and
I.~Kreykenbohm\inst{10,11}
}
\offprints{P.~Blay}
\institute{Institut de Ci\`encia dels Materials, Universitat de Val\`encia, PO Box 22085, 46071 Valencia, Spain\\
\email{pere.blay@uv.es}
\and
Departamento de F\'{\i}sica, Ingenier\'{\i}a de Sistemas y Teor\'{\i}a de la Se\~{n}al, Universidad de Alicante, Apdo. 99,
  03080 Alicante, Spain\\
              \email{ignacio@dfists.ua.es}
 \and
CRESST \& Department of Physics, University of Maryland Baltimore County, 1000 Hilltop Circle, Baltimore, MD 21250, USA
 \and
NASA Goddard Space Flight Center, Astrophysics Science Division, Code 661, Greenbelt, MD 20771, USA 
\and
Department of Physics and Santa Cruz Institute for Particle Physics, University of California, Santa Cruz, Santa Cruz, CA 95064, United States
\and
Massachusetts Institute of Technology, Kavli Institute for Astrophysics
and Space Research, Cambridge MA 02139, USA
  \and
IESL, Foundation for Research and Technology, 711 10 Heraklion, Crete, Greece 
\and
Physics Department, University of Crete, PO Box 2208, 710 03 Heraklion, Crete, Greece 
\and
 ESA, ESAC, P.O. Box 78, 28691 Villanueva de la Ca\~nada, Madrid, Spain
\and
Institut f\"ur Astronomie und Astrophysik, Abteilung Astronomie, Sand 1, 72076 T\"ubingen, Germany
\and
INTEGRAL Science Data Centre, 16 Ch. d'Ecogia, 1290 Versoix, Switzerland
}

\date{Received , ; accepted }
\abstract
   {In the past few years, a new class of High Mass X-Ray Binaries (HMXRB) has been claimed to exist, the Supergiant Fast X-ray Transients (SFXT). These are X-ray binary systems with a compact companion orbiting a supergiant star which show very short and bright outbursts in a series of activity periods overimposed on longer quiescent periods. Only very recently the first attempts to model the behaviour of these sources have been published, some of them within the framework of accretion from clumpy stellar winds.
   }
   {Our goal is to analyze the properties of \object{XTE~J1739-302}/\object{IGR~J17391-3021} within the context of the clumpy structure of the supergiant wind.
   }
   {We have used {\it INTEGRAL}~ and {\it RXTE}/PCA observations in order to obtain broad band (1\,--\,200 keV) spectra and light curves of \object{XTE~J1739-302} and investigate its X-ray spectrum and temporal variability.
   }
   {We have found that \object{XTE~J1739-302} follows a much more complex behaviour than expected. Far from presenting a regular variability pattern,   \object{XTE~J1739-302} shows periods of high, intermediate, and low flaring activity. 
   }
   {}
   \keywords{binaries: close, stars: supergiants, X-rays: binaries}
   \maketitle
%

\section{Introduction}

Wind-fed Supergiant X-Ray Binaries (SGXRBs) display high energy emission
arising from the accretion of material in the wind of an OB supergiant
onto the compact component of the system (a neutron star -NS- or black 
hole in orbit around the supergiant).  SGXRBs are persistent X-ray
sources, displaying an X-ray luminosity $L_{{\rm X}}\sim10^{36}\:{\rm erg}\,{\rm s}^{-1}$. Because 
of the physical characteristics of wind accretion, their emission is
variable on short timescales, with frequent flares, but relatively
stable on the long term (for example, the long-term
 {\it RXTE}/ASM lightcurve of Vela X-1, averaged and smoothed with a
 running window of 30\,d length, shows variations by only a factor of
 $\sim4$;  \citealt{ribo06}). If the orbit is eccentric, the
 luminosity is  modulated on the
orbital period of the system \citep[e.g.,][]{leahy02}. Stronger short
flares, with a fast rise and a typical timescale of the order of a few
hours, have been observed from several
systems, such as Vela X-1 \citep{lau95,krivonos03} or 4U~1907+09
\citep{fritz06}. 

Recently, thanks to the improved sensitivity of high energy missions,
many new SGXRBs have been discovered, leading to the suggestion of new
classes of X-ray sources. On the one hand, there is a number of
highly absorbed SGXRBs, invisible to previous
missions due to high absorption in the softer X-ray bands
\citep[e.g.,][]{chaty05}. On the other hand, Supergiant Fast X-ray Transients
(SFXTs) display fast outbursts, with a typical duration of a few hours, but
stay in quiescence most of the time \citep{smith06,neg06a,sgue06}. Unlike in   
classical SGXRBs, the X-ray luminosity of SFXTs goes down below the
sensitivity limit of the INTErnational Gamma-Ray Astrophysics Laboratory ({\it INTEGRAL}) and they 
remain undetectable for long time spans. 
They can only be observed
during an outburst or flare, for a short
time. Though several models have 
been proposed for this difference in behaviours, it seems to be a
natural consequence of the clumpy nature of OB star winds
\citep{walter07,neg08}. Sidoli et al. (\citeyear{sidoli07}) propose an alternative hypothesis, based on observations of
\object{IGR~J11215$-$5952}, in which the observed flaring activity is due to the interaction of the compact object with an extended equatorial
decretion disc around the supergiant star.

Although the definition of SFXTs as a putative new class of objects
was only possible when 
the optical counterparts to these systems started to be identified,
{\it INTEGRAL} has
contributed decisively to the characterization of the high energy
behaviour of these sources. So far, $\sim 12$ SFXTs or related objects
have been detected by {\it INTEGRAL} \citep{walter07,sgue06}. Among them, the
best characterized system is  \object{XTE J1739-302} = \object{IGR~J17391-3021},
generally taken as the prototype of the class \citep{smith06,neg06b}.

\object{XTE~J1739-302} was discovered by {\it RXTE} during a short 
outburst in 1997 \citep{smith98}, when it was detected only for a
period of a few hours. The source spectrum was well described by 
bremsstrahlung emission with a source temperature $kT\sim21$~keV. No
indications of any periodicity shorter than 300~s could be
found. During 2003, the source was detected by {\it INTEGRAL}/ISGRI
\citep{lutovinov05}. Again, a bremsstrahlung model with
$kT\sim22$~keV fitted the source spectrum well and no 
evidence of periodicity could be found.  A total of 6
outbursts were detected by {\it INTEGRAL} up to 2005
\citep{sgue05}. The mean duration of these outbursts is of the order 
of 5 hours and they are all highly structured. For a plot of {\it INTEGRAL}/ISGRI 
detections in the 20\,--\,40 keV energy range during the period 2003\,--\,2005, see Fig.~\ref{gps_lc}. 
The optical counterpart  was identified thanks to a {\it Chandra} localization as an O8\,Iab(f)
supergiant at a distance of $\approx2.3$~kpc \citep{neg06b}. 

In this work, we present a detailed analysis of {\it INTEGRAL} data
for \object{XTE~J1739-302} obtained mostly through the Galactic Plane Scans (GPS, public data), 
the Galactic Center Deep Exposure (GCDE, public data), and through three long exposures of the Galactic
Center, taken as part of the {\it INTEGRAL} Key Programme (KP)
observations. Section~\ref{sec:obs} will
be devoted to the description of the data and the 
analysis techniques used, including the presentation of 
results. An interpretation of these results in the context of our
current understanding of SFXTs and models of accretion from clumpy
winds \citep{walter07,neg08} will be presented in
Section~\ref{sec:disc}, followed by our conclusions.


\section{Observations, data reduction and analysis}
\label{sec:obs}

{\it INTEGRAL} is an ESA mission with contributions from Russia and
NASA (Winkler et al. \citeyear{winkler03}). There are 3 high energy instruments on board {\it
  INTEGRAL}. The SPectrometer on {\it INTEGRAL} (SPI, see
Vedrenne et al. \citeyear{vedrenne03}), the Imager on Board the {\it INTEGRAL} Satellite (IBIS, see
Ubertini et al. \citeyear{ubertini03}), and the Joint-European X-ray Monitor (JEM-X, Lund et al.
\citeyear{lund03}).  IBIS has two detector layers, ISGRI ({\it INTEGRAL}
Soft Gamma-Ray Imager, Lebrun et al. \citeyear{lebrun03}) and PICsIT (Pixellated Imaging Caesium Iodide Telescope, Labanti et al. \citeyear{labanti03}), operating in the 15--1000 keV and 1750 keV--10 MeV energy ranges respectively. Data from both layers
can be combined in the IBIS Compton mode \citep{lei97,forot07}. In this work we
make use of {\it INTEGRAL}/IBIS/ISGRI data and we will mean that when
we mention {\it INTEGRAL}/ISGRI for short. Data from the {\it INTEGRAL}/JEM-X instrument are sparse and only 
a few detections will be reported.

Exposures containing  \object{XTE~J1739-302} were obtained in 
 September\,--\,October 2006  and February\,--\,March 2007 (AO-4 Galactic
 Center Key Programme, from now on KP1, observations) and 
 August\,--\,September 2007 (AO-5 Galactic Center Key
 Programme, from now on KP2, observations).
The data analyzed correspond to {\it INTEGRAL} revolutions
478\,--\,481, 484, 485 (KP1, MJD 53990\,--\,54014), 534\,--\,537, 539\,--\,542 (KP1, MJD 54157\,--\,54184) and
594\,--\,596, 599\,--\,601, 604 (KP2, MJD 54337\,--\,54369). A revolution lasts for
$\sim3$ days and observations within a revolution are divided in exposures of
the order of $\sim2$~ks (called science windows). Publicly available
data mostly from GCDE observations during 2003-2005 have also been used. 

The data have been reduced with the standard Off-line Analysis Software
(OSA) version 6.0. OSA is distributed and maintained by the INTEGRAL
Science Data Center (ISDC\footnote{http://isdc.unige.ch}).

\subsection{Timing analysis}

\begin{figure*}
  \centering
 \resizebox{\textwidth}{!}{\includegraphics[height=0.2\textheight,width=0.7\textwidth]{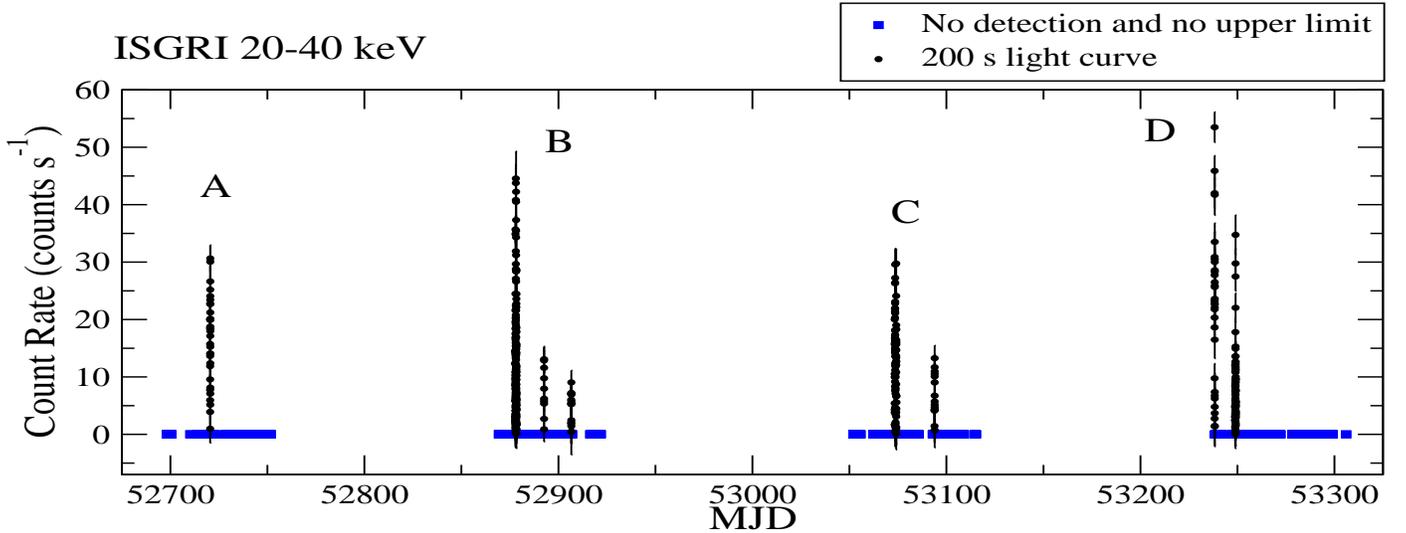}}
   \caption{Light curve of \object{XTE~J1739-302} data from {\it INTEGRAL}
     Galactic Center Deep Exposure (GCDE). In order to appreciate the number and
     structure of outbursts, a binning time of 200 s has been used.} 
              \label{gps_lc}%
    \end{figure*}

\begin{figure*}
 \centering
\resizebox{\textwidth}{!}{\includegraphics[width=1.0\textwidth]{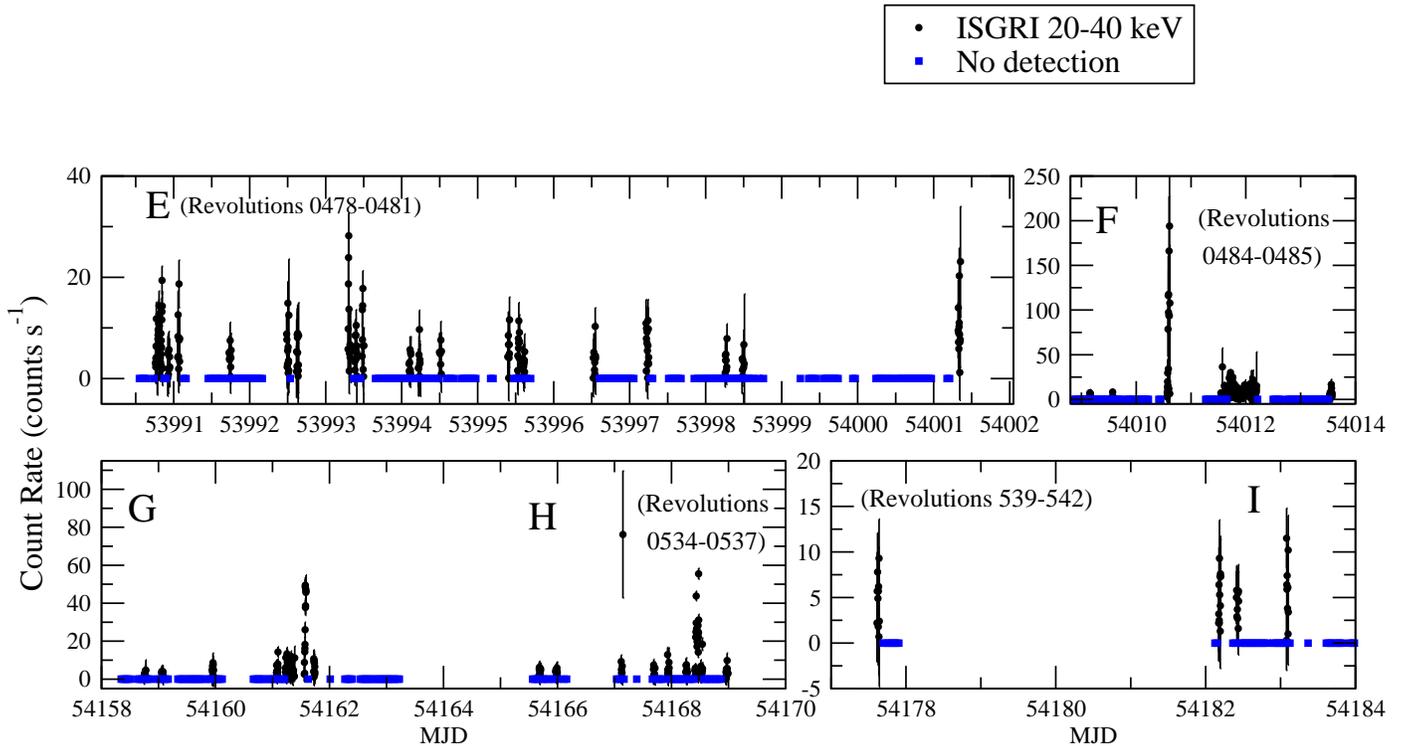}}
   \caption{Light curve of \object{XTE~J1739-302} data from {\it INTEGRAL} Key Programme 1 (KP1, September-November 2006 and February-March 2007). In order to appreciate the number and structure of outbursts, a binning time of 200 s has been used.}
              \label{kp1_lc}%
    \end{figure*}

The 20\,--\,40 keV lightcurve of \object{XTE~J1739-302} during GCDE
in the period 2003-2005, and during KP 1 are shown in Figs.
\ref{gps_lc} and \ref{kp1_lc}, respectively. During GCDE and KP1 observations there
is clear flaring activity. In contrast, during the KP2 period no
significant detection can be reported (but a 3-$\sigma$ upper limit on
the source flux of 5 count s$^{-1}$, that is $\sim43$~mCrab, can be
given\footnote{The Crab flux in the 20\,--\,40 keV energy band was obtained for 3 
pointings of revolution 300, namely pointings 39, 40 and 41, with OSA 6.0 software. 
A mean Crab flux of 117~count~s$^{-1}$ was used to 
convert the \object{XTE~J1739-302} flux from count~s$^{-1}$ to 
Crab units.}).  Times when {\it INTEGRAL} was staring towards the
source direction but the source is not detected are indicated in
Figs.  \ref{gps_lc} and \ref{kp1_lc} with zero flux (blue squares). 

JEM-X detections in two energy bands (4--15~keV and 15--30~keV) are summarized in Table \ref{table:jemx_setections}.
Due to its smaller Field Of View (FOV) the total 'on source' time is $\sim$100 times 
smaller than that of {\it INTEGRAL}/ISGRI. 

\begin{table}
\caption{Summary of {\it INTEGRAL}/JEM-X detections during GCDE and KP1 periods.}
\label{table:jemx_setections}
\begin{center}
\begin{tabular}{@{}cccc@{}}
\hline
 {\it INTEGRAL}  & MJD  & Count Rate       & Count Rate \\
     pointing    &      & 4--15 keV        & 15--30 keV \\
                 &      & (count s$^{-1}$) &  (count s$^{-1}$)  \\
\hline
\hline
00530065 & 52720.54 & 7.1$\pm$1.3 & 1.7$\pm$0.5 \\
00530066 & 52720.56 & 8.1$\pm$0.5 & 1.9$\pm$0.2 \\
01710057 & 53073.29 & 4.0$\pm$0.4 & 0.7$\pm$0.3 \\
01710067 & 53073.53 & 7.7$\pm$1.8 & 1.9$\pm$0.8 \\
01710077 & 53073.77 & 2.1$\pm$0.3 & 3.8$\pm$0.1 \\
\hline
04850027 & 54011.67 & 3.2$\pm$0.3 & 0.6$\pm$0.1 \\
04850036 & 54011.96 & 3.8$\pm$0.3 & 0.6$\pm$0.1 \\
05370061 & 54168.41 & 4.1$\pm$0.2 & 1.1$\pm$0.1 \\
05370062 & 54168.45 & 4.6$\pm$0.5 & 0.9$\pm$0.2 \\
\hline                                
\end{tabular}
\end{center}                   
\end{table}

During the GCDE period, the coverage is sparse and the lack of
homogeneity prevents the data from being useful for a statistically
significant determination of a characteristic outburst
frequency. There are three periods, however, during which the 
source was in the {\it INTEGRAL}/ISGRI FOV for a  time
interval long enough to ensure that there was no activity prior or after the
outburst activity. These periods are labelled A, B and C in
Fig.~\ref{gps_lc}. These activity periods have already been reported and
analyzed in \cite{lutovinov05}, \cite{sgue05}, \cite{sgue06}, and \cite{walter07}.

   \begin{figure} 
\centering
  \resizebox{\columnwidth}{!}{\includegraphics[width=0.42\textwidth]{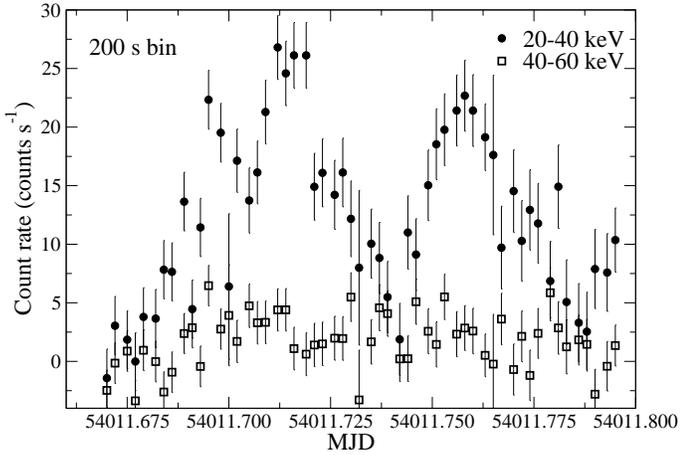}}
   \caption{Light curve of the four consecutive science windows (namely, 0027, 0028, 0029, and 0030) of revolution 485, the longest
set of consecutive observations in all the the KP1 period, with a total coverage of 8~ks. The time binning used for plotting purposes is 200 s.}. 
              \label{pwspe_rev485}%
    \end{figure}

Coverage during the KP1 period is much more homogeneous, as can be seen in
Fig.~\ref{kp1_lc}. We can define here 5
periods of continuous coverage, labelled E, F, G, H and I in
Fig. \ref{kp1_lc}. 
This light curve can be divided in 26 activity periods, following the definition used
by \cite{walter07}. In the majority of cases, these activity periods
correspond to a strong peak and some associated flaring activity, with secondary peaks before and after the maximum flux of
the peak. We call such an activity period an outburst. In a few cases, like the 4 consecutive science windows from revolution 458 shown in Fig. \ref{pwspe_rev485}, more than one strong peak fits within an activity period. The average duration of the activity periods is $\sim6$~hours, with a maximum duration of
16~h and a minimum duration of 0.6~h. We can differentiate two types of outbursts: those  with a mean count rate of $\sim$10~count~s$^{-1}$, fainter and more numerous; and the brighter ones, less frequent and with average flux of $\sim60$~count~s$^{-1}$, but reaching up to $\sim$190~count~s$^{-1}$. On average, both types of outbursts last for
 a very similar time span, but the brightest ones tend to last for longer than $\sim$1~h and the faintest can be as 
short as $\sim$0.6~h. This dataset can be used to search more consistently for a characteristic outburst
frequency and we will perform this task in Section~\ref{sec:freq}.

We have failed to detect any coherent modulation which could be associated with
the pulse period, up to time scales of the order of $\sim$1000 s. An apparent non-coherent 
modulation at about 4000~s is found in the KP1 data, see, e.g., the data for revolution 485 in \ref{pwspe_rev485}.
But this non-coherent modulation only reflects the characteristic time scale of the flaring activity during KP1, as 
described above.

\subsection{Spectral analysis}


Only those science windows with the source within the Fully
Coded Field Of View (FCFOV) of  
{\it INTEGRAL}/ISGRI and with a detection level above 7 for \object{XTE~J1739-302}
were considered for the spectral extraction\footnote{OSA software reports the ISGRI detections in a significance scale in
which a source is considered as detected if its fitted with a detection level of 7. This detection level of 7 will be equivalent to the classical
3$\sigma$ level above the measured noise.}. This
leaves us with a total of 32 spectra extracted for the public data in the 2003--2005
period and a total of 12 spectra for the KP 1
period. These spectra correspond to the brightest part of the
outbursts seen in Fig. \ref{gps_lc} and Fig. \ref{kp1_lc}. 
The spectra created per pointing have rather poor statistics and do not permit a detailed spectral
analysis. We have therefore created average spectra per revolution,
except for revolution 171 in which two spectra could be extracted
with sufficient statistics to test for spectral variability.


A bremsstrahlung model (bremss in XSPEC notation) was used in order to search for spectral variability and to
compare with previously published data \citep[e.g.,][]{smith98}. The
spectral parameters are shown in Table \ref{table:spe_bremss}. The fit
results are, within the uncertainties, all compatible
with a mean electron temperature of 21$\pm$2 keV and no spectral variation, in good
agreement with previously published values \citep[see][]{smith06,lutovinov05}. 

   \begin{table}
\begin{center}
      \caption[]{Spectral parameters for the bremsstrahlung and powerlaw models fit of {\it INTEGRAL}/ISGRI data. The given uncertainties are calculated for a 90\% confidence level.}
         \label{table:spe_bremss}
       \begin{tabular}{@{}l@{}ccccc@{}}
              \hline
Revolution   &  ${\rm kT}$                 &  $\chi^{2}_{{\rm red}}$/DOF & $\Gamma$  &  $\chi^{2}_{{\rm red}}$/DOF & Flux$\dagger$ \\
             & (keV)               &                             &           &                             & 20--100 keV \\  
\hline
\hline
0053  &  22  $^{+3}_{-2}$  &  0.8/20       & 3.10 $^{+0.16}_{-0.15}$   &  1.0/20 & 1.0 \\
0106  &  20  $^{+2}_{-1}$  &  1.6/20       & 3.17 $^{+0.11}_{-0.11}$   &  1.9/20 & 0.8 \\ 
0171a &  20  $^{+2}_{-2}$  &  1.2/20       & 3.14 $^{+0.15}_{-0.14}$   &  1.4/20 & 1.0 \\ 
0171b &  20  $^{+5}_{-4}$  &  0.7/15       & 3.2  $^{+0.3}_{-0.3}$     &  1.2/20 & 0.7 \\
\hline
0478  &  22  $^{+8}_{-5}$  & 0.8/20        & 3.1  $^{+0.4}_{-0.4}$   & 1.0/20  & 0.5 \\
0485  &  18  $^{+3}_{-2}$  &  0.8/20       & 3.28 $^{+0.20}_{-0.19}$   & 1.3/20  & 0.7 \\
0535  &  21  $^{+5}_{-3}$  &  1.3/20       & 3.04 $^{+0.25}_{-0.23}$   & 1.4/20  & 1.0 \\
0537  &  25  $^{+4}_{-3}$  &  1.6/20       & 2.89 $^{+0.16}_{-0.16}$    & 1.9/20  & 1.1 \\
   \hline

       \end{tabular}

\end{center}
$\dagger\times$10$^{-9}$ erg cm$^{-2}$ s$^{-1}$
  \end{table}

\begin{figure*}
 \centering
 \resizebox{0.85\textwidth}{!}{\includegraphics[]{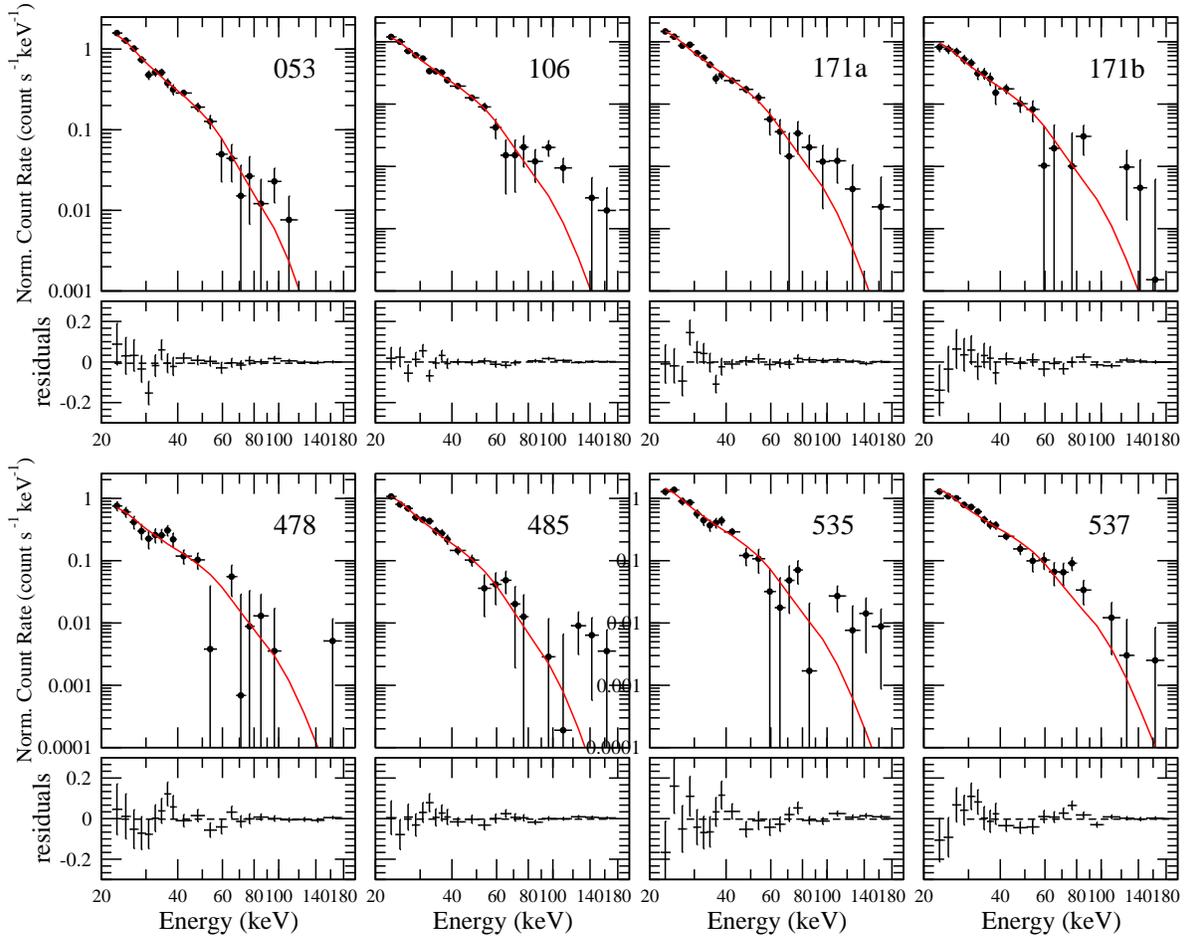}}
      \caption{Bremsstrahlung  fit to the
        average {\it INTEGRAL}/ISGRI spectra of revolutions shown in Table \ref{table:spe_bremss}. The corresponding revolution number is indicated in the top right corner of each plot. The fit shown is that of the bremsstrahlung model. Residuals are in units of count\,s$^{-1}$\,keV$^{-1}$.}
         \label{kp1_and_gps_spe}
\end{figure*}

Even though simple bremsstrahlung models produced
acceptable fits to the spectra, the residuals 
show the presence of possible absorption features at $\sim$30 keV and
$\sim$60 keV in some of the spectra (see Fig. \ref{kp1_and_gps_spe}).

A powerlaw model was also used to fit the spectra, for the sake of
comparison with other sources. The results are shown in Table
\ref{table:spe_bremss}. Apart from the possible absorption lines, which could 
not be seen in all spectra, no signs of spectral variability, within errors, could be found.

\begin{figure}
 \centering
 \resizebox{\columnwidth}{!}{\includegraphics[width=0.75\textwidth]{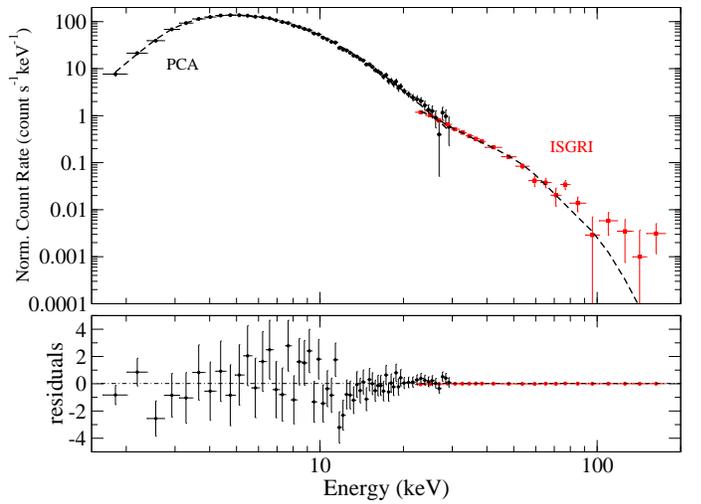}}
      \caption{Fit to the joint {\it RXTE}/PCA plus {\it INTEGRAL}/ISGRI average spectrum. Residuals are in units of count\,s$^{-1}$\,keV$^{-1}$.}
         \label{pca_spe}
\end{figure}

\begin{figure*}
 \centering
 \resizebox{0.99\textwidth}{!}{\includegraphics[]{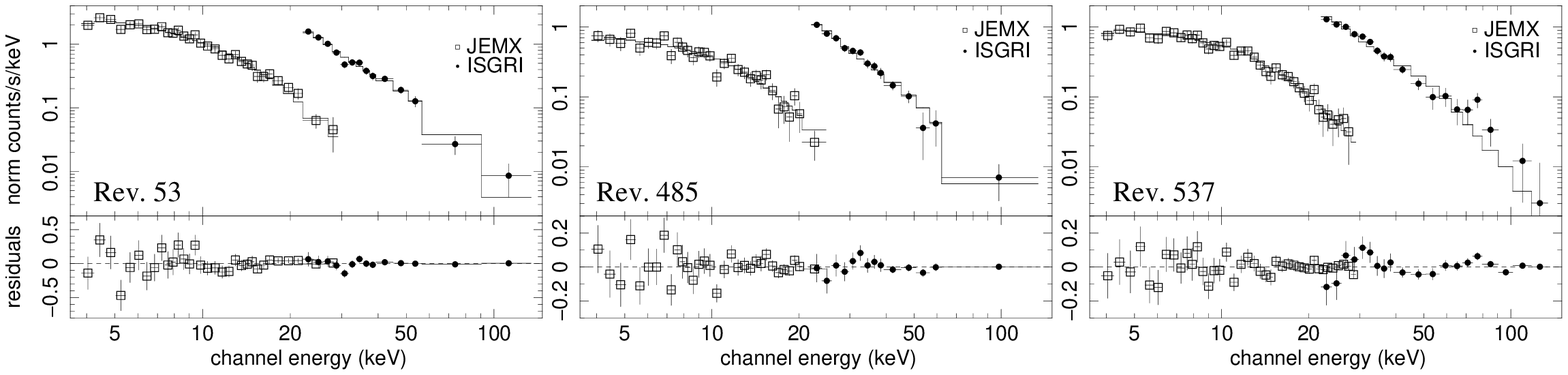}}
      \caption{Fit to the joint {\it INTEGRAL}/JEM-X plus {\it INTEGRAL}/ISGRI spectra from revolutions 53, 485, and 537. Residuals
     are in units of count\,s$^{-1}$\,keV$^{-1}$. }
         \label{jemx_spe}
\end{figure*}

With the aim of investigating the spectral behaviour below 20 keV, we have made use of
{\it RXTE}/PCA data (described in Smith et al. \citeyear{smith98}) and {\it INTEGRAL}/JEM-X spectra from revolutions 53, 485 and 537.
A spectrum from {\it INTEGRAL}/JEM-X data could also be extracted for revolution 171, but it was very noisy and not useful for spectral fitting. 
The {\it RXTE}/PCA spectrum together with the GCDE {\it INTEGRAL}/ISGRI average spectrum can be seen in Fig. \ref{pca_spe}. All joint {\it INTEGRAL}/JEM-X\,--\,{\it INTEGRAL}/ISGRI spectra can be 
seen in Fig. \ref{jemx_spe}.  A powerlaw, modified by a photo-absorption at low energies plus a cut-off at high energies, was used as the basic model and the photo-absorption column was fixed to 4.2$\times$10$^{22}$~cm$^{-2}$ \citep{smith98} in all cases.

 A summary of the fit of these joint PCA/JEM-X\,--\,ISGRI spectra is shown in Table \ref{table:joint_spe}. 

\begin{table*}
 
\begin{center}
\caption{Summary of best fit parameters obtained when 
joining {\it RXTE}/PCA data together with the GCDE {\it INTEGRAL}/ISGRI average spectrum,
and {\it INTEGRAL}/ISGRI and {\it INTEGRAL}/JEM-X from revolutions 53, 485, and 537. Uncertainties and upper limits are
given at a 90\% confidence level. The photoabsorption column was fixed to 4.2$\times$10$^{22}$~cm$^{-2}$ \citep{smith98}. 
     The cross-calibration factor was set to unity for {\it RXTE}/PCA and for {\it INTEGRAL}/JEM-X spectra.}
\label{table:joint_spe}
\begin{tabular}{@{}cccccccccc@{}}
\hline
Joint   &  N$_{\rm H}$                 & Cross-Calibration & $\Gamma$   &  $E_{\rm cut}$   &  $E_{\rm fold}$  &  $\chi^2_{red}$/DOF   \\
Spectra &  $\times$10$^{22}$ cm$^{-1}$ & Constant          &            & (keV)            & (keV)            & \\
\hline
\hline
PCA+ISGRI$_{AV}$           & 4.2 & 0.59$\pm$0.03  & 1.24$\pm$0.04  & 6.2$\pm$0.5  & 18$\pm$1 & 1.0/79   \\
\hline
Rev. 053, JEM-X+ISGRI   & 4.2  & 0.80$\pm$0.10  &   1.80$\pm$0.11   &  $<$11 & 29$^{+6}_{-5}$  &  1.00/56 \\
Rev. 485, JEM-X+ISGRI   & 4.2  & 1.8$\pm$0.3    &   1.7$\pm$0.2     &  $<$11 & 22$^{+6}_{-4}$  &  1.03/42     \\
Rev. 537, JEM-X+ISGRI   & 4.2  & 1.3$\pm$0.2    &   1.56$\pm$0.11    &  $<$11 & 28$^{+6}_{-5}$ &  1.10/56 \\
\hline
\end{tabular}
\end{center}

\end{table*}

Regardless of the continuum models used, for the datasets of
revolutions 53, 106, and 478, we see residual features reminiscent of
broad absorption structures at $\sim$30~keV  for the former, $\sim$65~keV  for revolution 106 and
$\sim$30~keV and $\sim$60~keV for the latter (see Fig. \ref{kp1_and_gps_spe}). It is tempting to associate
them with a Cyclotron Resonant Scattering Feature (CRSF) and its first harmonic, but adding such features
to the spectral model does not lead to a significant improvement
of the fit. A deeper exposure of the source would be required
in order to get sufficient statistics for an analysis of these
possible spectral features.


\section{Discussion}
\label{sec:disc}


\subsection{Outburst frequency}
\label{sec:freq}

\citet{walter07} attempt to define a typical
flare duration in SFXTs, in order to quantify the frequency of
outbursts. They identify a typical flare duration of 3~h, though
flares are sometimes seen to be part of longer flaring
episodes.\citet{walter07} define a flaring episode as separated by at least
25~ks of {\bf non-detection} from another episode. During the time span  
2003\,--\,2005, they count a total of 13 of these flaring episodes from 
\object{XTE~J1739$-$302} detected during a total observed elapsed time
of 126.4\,d. Of these, only one can be interpreted as a longer ($\sim
14$~h) flare, but it is also possible to interpret it as a superposition of three shorter
flares. The rest of the flares last between $\sim2$ and $\sim8$~h,
with a typical duration of 4.2~h, similar to other SFXTs. In any case,
due to the non-continuity of the observations it 
is difficult to set a characteristic outburst frequency for this
period.

\begin{table}
\begin{center}
\caption{Frequency of outbursts for the activity periods defined
  during KP1 observations.}  
\label{table:kp1_outb_count}
\begin{tabular}{@{}lcccc@{}}
\hline
 Period & Interval & On source & Num. of  & Outburst \\
        &          & time      & outburst & Frequency  \\
        & (MJD)    & (d)       &          & (d$^{-1}$)  \\
\hline
\hline
E       & 53990.50\,--\,54001.50   & 5.5 & 21   & 3.8 \\
F       & 54008.75\,--\,54014.00   & 3.1 & 13    & 4.2 \\
G       & 54158.00\,--\,54163.25   & 3.0 & 8    & 2.7 \\ 
H       & 54165.50\,--\,54169.00   & 1.8 & 8    & 4.4 \\
I       & 54182.00\,--\,54184.00   & 1.3 & 3    & 2.3 \\
\hline
\end{tabular}
\end{center}
\end{table}
%

 Data during the KP1 observations offer a much more homogeneous
 coverage. 
 Table~\ref{table:kp1_outb_count} summarizes the outburst counting,
 showing the duration of the quasi-continuous observing period,
the number of outbursts, as seen in Fig.~\ref{kp1_lc}, and the
derived outburst frequency for each period. This is obtained by simply
dividing the number of outbursts seen by the time duration of the observing
period (considering only the 'on-source' time). 

The results are surprising, because the outburst frequency derived is
much higher than previously expected (2\,--\,4~outbursts~d$^{-1}$ as
opposed to 0.14~outbursts~d$^{-1}$ given as average of {\it INTEGRAL}
detections of SFXTs by \citealt{walter07}) and
does not seem to vary much between observations. This higher frequency
is also seen in the active periods during the GCDE observations, such
as the MJD 52877\,--\,78 and 53073\,--\,74 intervals in 
Fig.~\ref{gps_lc}. 

The average frequency for the KP1 observations is 3.5~outburst~d$^{-1}$. 
However, no outbursts have been seen during the whole
KP2 observations. Examination of Fig.~\ref{gps_lc} shows that long
time intervals are characterized by low X-ray activity. We know that
the absence of activity is not complete, as at least one outburst from 
\object{XTE~J1739$-$302} was observed by the {\it RXTE}/PCA during its
Galactic Bulge Scan observations on 54347.5 MJD. 
 This outburst took place during the long gap in the I bin of KP2 observations\footnote{Craig Markwardt, private communication}.

\begin{figure*}
   \centering
   \resizebox{\textwidth}{!}{\includegraphics[angle=0,width=\columnwidth]{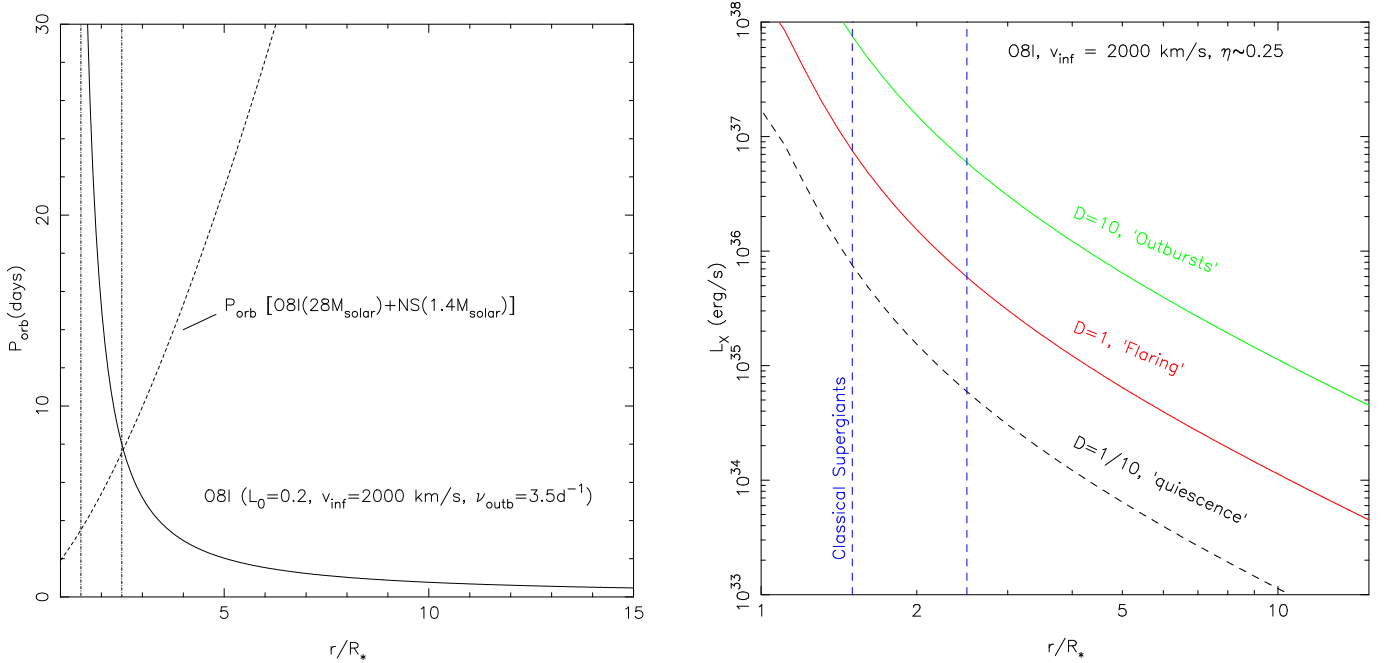}}
   \caption{{\bf Left panel:} we show the orbital periods corresponding to a
   certain outburst frequency (3.5 per day, in this case) versus the
   radial distance (solid line). The dashed line represent
   the Kepler law for a NS orbiting an O8I star. Vertical lines mark the smooth
   wind regime region, where the classical supergiants reside. For the
   observed frequency, the NS must lie close to the smooth region
   upper limit, at $r\lesssim 3R_{*}$. {\bf Right panel:} accretion
   luminosity versus radial distance. At the distance to the system,
   the luminosity is $\sim 5\times 10^{35}$ erg s$^{-1}$, as observed
   in the flares. Relative changes in the density by a factor of 10
   (plus or minus) can led to outburst or 'quiescence'.   } 
              \label{fig:lxperiod}
\end{figure*}


Taking the average outburst frequency for KP1 at face value\footnote{If KP2 was carried out during an unusually inactive
  period of this source while KP1 occurred during an unusually high
  activity period, it would make sense to consider an \emph{average}
  outburst frequency of $\nu_{\rm outb}\approx 1.5$ d$^{-1}$. Then,
  maintaining \object{XTE~J1739-302} outside the smooth wind zone of Fig. \ref{fig:lxperiod} (left panel) would
  require $L_{0}\approx 0.25$. The values of $r$ and $P_{\rm orb}$ do
  not change too much and the terms of the discussion are nearly the
  same.}, we can compute the
orbital period of a compact object necessary to match the observed
frequency, as a function of the radial distance. This can be
accomplished by taking into account that  
 
\begin{equation}
\nu_{\rm outb}=\frac{N(r)_{\rm c}^{{\rm ring}}}{P_{\rm orb}}
\end{equation}

where $N_{\rm c}^{{\rm ring}}$ is the number of clumps inside a ring at a
distance $r$ from the donor star. This quantity 
represents the number of
clumps that a NS orbiting in a circular orbit of radius $r$ will
\emph{statistically} be able to 'see'  \citep[and accrete, see][]{neg08}. The $P_{\rm orb}$ vs. $\nu_{\rm outb}$ 
relationship is depicted in
the left panel of Fig. \ref{fig:lxperiod} by the solid line.

On the other hand, we can calculate the orbital period of a binary system
by simply applying Kepler's third law, and express it as a function of
binary separation. This has been depicted as a dashed line in
Fig. \ref{fig:lxperiod} for a NS of canonical mass $1.4\:M_{\sun}$
orbiting a O8\,I star.

For the high frequency observed, the clumping parameter $L_{0}$
\citep{oskinova,neg08} must have a value of $\sim0.2$, close to the lower bound 
of the range suggested by \citet{oskinova} 
(0.2\,--\,0.5) based on independent optical and ultraviolet observations. For higher values
of $L_{0}$ the system would lie in the smooth wind regime. 

In Negueruela et al. (\citeyear{neg08}) we stated that a value of $L_{0}\approx0.35$
was adequate outside the smooth regime zone to reproduce the observed
properties of some fast transient systems. However, we argued, its
value must be lower ($L_{0}\approx0.1$) inside the asymptotic zone in
order to produce the persistent emission seen in the classical
systems. One possibility to circumvent this discrepancy was to
consider a radial dependence of this parameter, increasing
outwards. If this is the case, it is entirely natural that, in the
(narrow) transition zone -from the smooth wind regime to the fully
developed clumps- where XTE~J1739$-$302 orbits, presents an
intermediate value. 

Under this interpretation, the system should have therefore an orbital
separation of $r\lesssim 3R_{*}$ and a $P_{\rm orb}\approx 8$ d. The
proximity of the NS to the smooth regime zone is consistent with the
high activity displayed by the source.

Inspection of Fig. \ref{kp1_lc} shows that the vast
majority of outbursts have an average maximum flux of $\sim 10-12$
counts~s$^{-1}$. Superimposed to this 'flaring activity' (faintest outbursts) there 
are outbursts with maximum flux several times higher than this flaring
level. For example, in bin E, we have two outbursts that surpass
the 20 counts~s$^{-1}$ barrier. In bin F, we have one with an outstanding
$\sim 200$ counts~s$^{-1}$. In bins G and H, we have two outbursts
reaching $\sim 50-60$ counts~s$^{-1}$. The flaring in bin I is again
normal.

The distance to this source is $d=2.3$ kpc (Negueruela et
al. \citeyear{neg06b}). Therefore these fluxes translate into the following
luminosities: the flaring activity is at a level of $\sim 5 \times
10^{35}$erg~s$^{-1}$. The two outbursts in bin E reach luminosities
of $\sim 10^{36}$~erg~s$^{-1}$ and those in bins G and H $\sim 2.5
\times  10^{36}$~erg~s$^{-1}$. Finally, the strong outburst in bin F
reaches $\sim 10^{37}$~erg~s$^{-1}$. In other words, the bright outbursts are
between 2 and 20 times more powerful than the faint ones. Otherwise, they
seem to be identical. Interestingly enough, the bright outbursts seem to be
regularly separated by an interval of $\sim 8$ d (MJDs $\sim$53993.5,
$\sim$54001.4,$\sim$540010.5,$\sim$54160.7,$\sim$54168.5), which coincides with
the orbital period deduced from our model.\footnote{And also with the
outburst frequency found by Walter and Zurita Heras (\citeyear{walter07}). These
authors were detecting only the strong outbursts.}  In the following lines, and 
with the aims to make the text clearer, with outburst we will mean only the brightest ones, and will refer
to the rest as 'flares' or 'flaring activity' periods.

To investigate whether these luminosities can fit into our scenario, we
have plotted the accretion luminosity of simple Bondi-Hoyle wind accretion,
following the formalism described in \cite{reig03}. This is plotted
in the right panel of Fig. \ref{fig:lxperiod}. We have chosen an efficiency factor of
$\eta=0.25$ appropriate for accretion on to a NS. Furthermore, we have
applied a multiplicative factor adjusted to match the range of
observed luminosities in classical supergiants (namely, $\left[10^{36} -
  10^{37}\right]$~erg~s$^{-1}$), inside the smooth wind zone. At
$r=2\,R_{*}$ we have 
$L_{{\rm X}}=2 \times 10^{36}$~erg~s$^{-1}$, as observed in the O8-9\,Ia star
4U~1907$+$09 \citep{cox05}. We also show the predicted luminosities increased or
decreased by a factor of 10. Now, assuming that the putative NS is at
$r\lesssim 3R_{*}$, we can see that the luminosities predicted by
this model agree very well with those observed in the flaring
activity. In other words, the majority of clumps have the same density
and velocity predicted by the smooth wind approximation at this
distance. In order to produce the outbursts we only need to invoke
increments in the density of the clumps by the corresponding factors 2
to 20.  On the other hand, a decrease in density of about the same factor,
would bring the source under the level of detectability. Therefore,
density contrasts of $\sim 10^{1}-10^{2}$ would explain all but the
strongest outburst. Such contrasts are lower than expected in
most formulations of the porous wind model \citep[e.g.,][]{runacres02,owocki06}.

During KP2, no outbursts or flaring activity were detected. We are able
to set a 3-$\sigma$ upper limit on the flux of 43~mCrab. This
corresponds to a luminosity of $\sim 2.5 \times 10^{35}$ erg
s$^{-1}$. Since this upper limit in KP2 is close to the flaring level in KP1,
the lack  of flaring detection could not mean, necessarily, a real
lack of activity (in fact, the upper limit found for the luminosity is well above
the typical quiescence luminosity reported by Sidoli et al. \citeyear{sidoli08}).
A drop in the density of the wind by only a factor
of 2.5 could bring the flares already under the level of
detectability. In fact, as mentioned above, \emph{RXTE} observed a flare
during the long gap in the KP2 observations.

The lack of outbursts is more complicated to 
explain. While the lack of any detection during KP2 would still be
consistent with the presence of a low luminosity flaring (induced by a
decrease in wind density, for example), the lack of outbursts is real.

We have seen before, that these outbursts seem to be separated by
$\sim 8$\,d, very similar to the predicted orbital period. In
principle, the most straightforward explanation is that these are
produced at or close to the periastron passage on an eccentric
orbit. In order to reproduce the ratio of luminosities
betwen flares and the outbursts (a factor 2 to 5, not considering the
large outburst of bin F) would require a degree of eccentricity ~0.1.
However, there is no special reason why we should observe 
no outbursts during KP2.


%
%
%
%

\subsection{Lack of pulsations}



In principle, the possibility that the compact object in \object{XTE J1739-302} is a black hole can not be completely discarded. However, due the photon index found in the spectral fit and the marginal evidence of a possible CRSF we will consider in the following discussion that the compact component in \object{XTE J1739-302} is a NS. If this is the case, the absence of pulsations can be explained by geometrical effects. This possibility has also been considered in order to explain the lack of pulsations in \object{4U~1700-37} \citep{white83}, \object{4U 1700+24} \citep{masetti02}, and \object{4U~2206+54} \citep{blay05}. 
Two scenarios can be invoked: a) in the first one, the magnetic axis and the spin axis of the NS are almost aligned, in this configuration only one pole of the NS  is continuously visible; b) the second scenario would imply that the orbit of the binary system has a very small inclination angle.

Furthermore, the possibility of very slow pulsations cannot be excluded. Although our data set is not suitable for a proper timing analysis, because of the many gaps between the observations, we have not found evidence of pulsations on time scales up to 1000~s. \cite{smith98} discarded the presence of pulsations below 300~s. \object{2S~0114+650}, an X-ray pulsar accreting matter from the wind of a supergiant B1 companion (Reig et al. \citeyear{reig96}), shows pulsations with a spin period of 2.8~h \citep{finley94}. Another example of a neutron star showing very slow pulsations is the case of that in the Low Mass X-Ray Binary System \object{4U~1954+319}, which pulsates with a period of 5 hours \citep{mattana06,corbet08}. The inhomogeneity of our data at this time scale hinders the detection of a modulation of this order of magnitude.

We have considered that the X-ray variability directly traces the supergiant wind behaviour, however the accretion process could be more complex and, then, our conclusions would not be derived in such a straightforward manner.
Grebenev \& Sunyaev (\citeyear{grebenev07}) suggest that the flaring behaviour of fast X-ray transients can be due to the onset and offset of the propeller effect in those systems with a NS companion. This would happen in systems in which the NS possesses a spin period close to a certain limiting value $P_{prop}$. However if the absoption features suggested by our data at the energies of $\sim$30~keV and $\sim$60~keV correspond to a real CRSF and its first harmonic, that would imply a magnetic field strength of the order of $\sim$3$\times$10$^{12}$~G (according to the formula $[B/10^{12} \rm{G}]=[E/\rm{keV}](1+z)/11.6$,  which relates the energy of the CRSF, E, with the magnetic field strength, B, and where $z$ represents the gravitational red-shift at which we see the absorbing region). This value is well within the range 1.3\,--\,4.8$\times10^{12}$~$\rm{G}$ for the expected magnetic field strength of a NS obtained by \cite{coburn02}. The NS in \object{XTE~J1739-302} does not seem to rotate at a particularly fast rate (periods below 1000~s are discarded) and possibly shows a typical value for its magnetic field strength. Thus, we do not find reasons why the accretion process in this system could be different from the majority of accreting systems and why the propeller effect could be present.  Furthermore, if we take the typical values of $m_{NS}$=1.4M$_{\sun}$, $R_{NS}$=10~km, $v_{NS}$=10$^3$km~s$^{-1}$ for the mass, radius and relative velocity to the stellar wind of the neutron star, and we take into account the estimated possible magnetic field of  $\sim$3$\times$10$^{12}$~G  and the possible 8~d orbital period, we obtain, according to the formalism of \cite{grebenev07}, that only for spin periods faster than $\sim$10~s the propeller effect would take place, for mass loss rates from the donor star on the order of
$\sim$10$^{-5}$M$_{\sun}$~yr$^{-1}$. The possibility to reach this limiting $P_{prop}$ is ruled out by the  fact that pulse periods shorter than $\sim$1000~s are discarded. It must be noticed that the derivation of this limiting value is done in the framework of the smooth wind approximation and, therefore, may have to be re-formulated in order to account for the peculiarities of the clumpy wind formalism.

\cite{titarchuk02} proposed the smearing out of pulsations due to electron scattering in an optical thick environment. This model was already considered as a possible explanation for the lack of pulsations in the HMXRB \object{4U 2206+54} by \cite{torrejon04}. Due to the high absorption column found in \object{XTE 1739-302} \citep{smith98}, the possibility of this mechanism to be operating in this system should not be discarded.

\section{Conclusions}

\object{XTE~J1739-302}  has been observed with the {\it INTEGRAL} observatory during a long time span. In the period 2003\,--\,2005, as part of the {\it INTEGRAL} core program observations of the Galactic Center, the source showed moderate activity, with an average outburst frequency below 1~outburst~per day. During the deep exposures of the Galactic Center taken within the frame of our first Key Programme observations, in the 2006\,--\,2007 period, \object{XTE~J1739-302} showed a higher level of activity, with a mean outburst frequency of $\sim$3~outburst~per day. Surprisingly, during the last observations in 2007 (within the second run of our Key Programme) the source
showed an unusually low activity state, and no outburst was detected with a flux above 43~mCrab. The behaviour during the first two periods can be explained well within the framework of the clumpy wind models. These models will not only explain the observed properties of the source but will also predict an orbital periodicity around $\sim$8~d and suggests the presence of a NS as the compact companion. To explain the low activity observed during KP2 period, geometrical considerations related to the eccentricity of the orbit, or a drop in the mass loss from the supergiant companion, need to be invoked in order to maintain consistency with the proposed model. A continuous monitoring of the system will allow disentangling which of the three observed behaviours (non dectability, moderate activity, and high activity) is representing the {\it normal} state of the source, will lead to the careful determination of the model parameters, and will permit to constrain the geometrical parameters of the system.

An independent constraint or measure of the orbital period of \object{XTE~J1739-302} would help to support or discard the clumpy wind model as the explanation to the behaviour of \object{XTE~J1739-302}. This model would, then, be the first consistent attempt to explain the observational properties of the SFXT and the classical wind-fed supergiant systems all together. The long spans of high activity
seem incompatible with the model proposed by Sidoli et al. (2007), where
outbursts happen once or twice each orbital cycle. Moreover, the
observations presented here represent strong evidence against a coherent
periodicity in the recurrence of the outbursts, which is a requirement of
the model. It must be stressed that this model was specifically designed
for \object{IGR~J11215$-$5952}, a system presenting periodic outbursts, and its
applicability to SFXTs is just a hypothesis, even if it turns to be
appropriate for this particular source.

\begin{acknowledgements}

We are grateful to Craig Markwardt for sharing RXTE data during
INTEGRAL KP2 period.\\
SMN is a researcher of the Programme
{\em Juan de la Cierva}, funded by the Spanish Ministerio de Educaci\'on y
Ciencia (MEC) and the University of Alicante, with partial 
support from the Generalitat Valenciana and the European Regional
Development Fund (ERDF/FEDER).
This research is partially supported by the MEC under
grants AYA2005-00095 and CSD2006-70. \\
JMT aknowledges the support by the Spanish Ministerio de Educaci\'on y
Ciencia (MEC) under grant PR2007-0176.\\
PR acknowledges the support by the European
Union Marie Curie grant MTKD-CT-2006-039965. \\
This research has made use of data obtained
through the INTEGRAL Science Data Center (ISDC), Versoix,
Switzerland.
 
\end{acknowledgements}

\end{document}